\documentclass[12pt]{aastex}
\usepackage{emulateapj5}
\usepackage{natbib}
\usepackage{lscape}
\usepackage{multicol}
\usepackage{amsmath,lipsum}
\usepackage{longtable}

\slugcomment{Version from \today}

\usepackage{graphicx}
\usepackage{subfigure}
\usepackage{lscape}

\begin{document}

\title{Hot Core, Outflows and Magnetic Fields in W43-MM1 (G30.79 FIR 10)}

\author{T. K. Sridharan\altaffilmark{1}, R. Rao\altaffilmark{2}, K. Qiu\altaffilmark{1,3,4}, P.
Cortes\altaffilmark{5}, H.  Li\altaffilmark{1,6,7}, T. Pillai\altaffilmark{1,8}, N. A. Patel\altaffilmark{1},
Q. Zhang\altaffilmark{1}}
\altaffiltext{1}{Harvard-Smithsonian Center for Astrophysics, 60 Garden Street, Cambridge, MA 02138, USA}
\altaffiltext{2}{Submillimeter Array, Academia Sinica Institute of Astronomy and Astrophysics, 645 N.
Aohoku Place, Hilo, HI 96720, USA}
\altaffiltext{3}{Max Planck Institute for Radioastronomy, Bonn 53121, Germany}
\altaffiltext{4}{School of Astronomy and Space Science, Nanjing University, Nanjing 210093, China}
\altaffiltext{5}{National Radio Astronomy Observatory-Joint ALMA Office, Alonso de Cordova 3107, Vitacura,
Santiago, Chile}
\altaffiltext{6}{Max Planck Institute for Astronomy, Hiedelberg, Germany}
\altaffiltext{7}{Dept. of Physics, The Chinese University of Hong Kong, Hong Kong}
\altaffiltext{8}{Astronomy Department, California Institute of Technology, 1200 East California Blvd., Pasadena, CA 91125, USA}
\altaffiltext{9}{email: tksridha@cfa.harvard.edu}


\begin{abstract}
We present submillimeter spectral line and dust continuum polarization observations of a remarkable hot core 
and multiple outflows in the high-mass star-forming region W43-MM1 (G30.79 FIR 10), obtained using the 
Submillimeter Array (SMA). A temperature of $\sim$ 400 K is estimated for the hot-core using CH$_3$CN (J=19-18) 
lines, with detections of 11 K-ladder components.  The high temperature and the mass estimates for the outflows 
indicate  high-mass star-formation. The continuum polarization pattern shows an ordered distribution, and its 
orientation over the main outflow appears aligned to the outflow. The derived magnetic field indicates slightly 
super-critical conditions. While the magnetic and outflow energies are comparable, the B-field orientation 
appears to have changed from parsec scales to $\sim$ 0.1 pc scales during the core/star-formation process.
  
\end{abstract}


\maketitle
\section {Introduction}
The process of high-mass star-formation and early evolution is marked by the phenomena of outflows (for example 
Shepherd \& Churchwell, 1996; Beuther et al. 2002; Zhang et al. 2005) and hot cores (e.g Cesaroni, Walmsley 
\& Churchwell, 1992;  Cesaroni et al. 2010).  Precious little is known about the role of magnetic fields in 
the process (e.g. DR21(OH), Lai et al 2003; Girart et al. 2013; G34.41 Cortes et al., 2008; 
G31.41, Girart et al., 2009; IRAS18089-1732, Beuther et al, 2010; W51, G5.89,  Tang et al., 2009a,b,2013). 
Due to the large distances to high-mass star-forming 
regions, interferometric studies are critical to gain better understanding, as in the above examples. It is 
also necessary to study regions in early stages of evolution, preferably the high-mass proto-stellar object 
phase (HMPOs, e.g. Sridharan et al 2012) or younger, to probe pristine conditions. In this Letter, we report 
the discovery of a high temperature hot-core and multiple outflows towards such an early stage object, 
W43-MM1 (G30.79 FIR10), and the magnetic field distribution around it, using submillimeter wavelength 
spectral line and continuum polarization observations at 345 GHz, obtained with the Submillimeter Array\footnote{The Submillimeter Array is a joint project between the Smithsonian
Astrophysical Observatory and the Academia Sinica Institute of Astronomy and Astrophysics, and is funded
by the Smithsonian Institution and the Academia Sinica.}(SMA; Ho, Moran \& Lo, 2004).

\section {Object and Observations}

W43-MM1, also known as G30.79 FIR 10, is the brightest dust emission core in the W43 "mini-starburst" 
region (Motte, Schilke \& Liz, 2003; Bally et al 2010). The region harbours UC-H{\sc ii} regions and water 
and methanol masers marking high-mass star-formation.  The  FIR10/MM1 core is at the head of a cometary 
infra-red dark cloud located at 5.5 kpc (from maser parallax; Zhang et al, 2013), facing a giant H{\sc ii} 
region powered by a cluster of WR-OB stars. It is devoid of cm-wavelength emission which may suggest an early, 
HMPO evolutionary stage. In addition, evidence for infall in multiple spectral lines, accretion at high rates and extended SiO emission possibly 
due to colliding flows have also reported (Cortes et al., 2010; Cortes, 2011; Herpin et al., 2012; Nguyen-Luong et al, 2013),
making the region very interesting.
Polarized dust emission at 1.3mm suggested an inconclusive pinched morphology for the magnetic field (Cortes \&
Crutcher, 2006), with a field strength of 1.7 mG, implying a near
critical mass-to-magetic flux ratio of 0.9, later refined to 1.9 (Cortes, 2011). If magnetic support was
dominant in the natal core, a pinched morphology is expected.  

The SMA observations
reported here, intended to delineate the B-field better, and to study line emission from the core, were 
obtained in the sub-compact and compact configurations. The observations were conducted on
24 April, 2007 and 28 May, 2008  under excellent weather  conditions with a 225 GHz zenith opacity of 
$\sim$ 0.05. The sub-compact observations only had 5 antennas.  The receivers were tuned to
346.51 \& 349.42 GHz at the center of the upper side-band for the two observations with the phase center
at ($\alpha,\delta$) = 18:47:47.0,  -01:54:30.0 (J2000). The correalator was configured to provide a uniform 
resolution of 0.81 MHz (0.7 kms$^{-1}$ at 349 GHz).  The combined data have a UV coverage of 16-160 $k\lambda$.  The 
polarization, phase and flux calibrators were 3c273, 1791-09 and Uranus respectively. Standard data reduction 
procedures were used under MIRIAD.  A discussion of the SMA polarimetry system can be found in Marrone et al. (2006) 
and Marrone \& Rao (2008). For continuum, multi-frequency synthesis was used to combine the two data sets with 
slightly differing frequencies, by 3 GHz, for the two tracks.

\section {Continuum Emission and Magnetic Field}

The continuum emission, mapped using the combined data from the two configurations at a beam size of 2.5$''\times$2.1$''$, shows 
multiple peaks (fig. 1) with 
integrated fluxes and masses in the range 1$-$10 Jy and 100$-$1000 M$_\odot$ (Table 1 ). All the parameters of
the cores were obtained by gaussian fitting and deconvolution. The masses were 
estimated following Hildebrand (1983), using a dust absorption coefficient of 2 cm$^2$g$^{-1}$ at 343 GHz
(Ossenkopf \& Henning 1994), a gas-to-dust ratio of 100 and a temperature of 25 K from SED 
fitting (Bally et al 2010) for all cores except the hot core A (see section 4), where a range of 25K - 300 K was used.  

\begin{table*}
    \begin{tabular}{ccccccccccc}
      \multicolumn{10}{c}{Table 1: Continuum Emission}\\
      \hline
      \hline

ID&    RA&        DEC& $\Delta$RA, $\Delta$DEC &      Peak, Err& Intg.& Maj& Min& PA&      Mass \\
&  (J2000.0) & (J2000.0) & $\prime\prime$ & Jy/bm  & Jy & $\prime\prime$ & $\prime\prime$ &  K & M$_{\odot}$ \\
\hline
A & 18:47:47.00  & -1:54:26.6 & 0.1 0.1 & 7.9,  0.61 & 11.5 & 1.7 & 1.3 & -52 & 920 - 80 \\
B & 18:47:46.86  & -1:54:29.7 & 0.2 0.2 & 3.5,  0.52 &  6.2 & 2.1 & 1.9 & -63 & 500 \\
C & 18:47:46.41  & -1:54:32.9 & 0.2 0.2 & 1.9,  0.26 &  3.7 & 2.7 & 1.7 &  63 & 300 \\
D & 18:47:46.69  & -1:54:32.3 & 0.2 0.2 & 1.1,  0.15 &  1.9 & 2.1 & 1.5 & -69 & 150 \\
E & 18:47:47.05  & -1:54:31.0 & 0.2 0.2 & 1.2,  0.16 &  1.6 & 1.7 & 0.9 &  28 & 130 \\
F & 18:47:46.55  & -1:54:23.1 & 0.1 0.2 & 0.7,  0.08 &  1.4 & 3.4 & 0.9 &  22 & 110 \\
      \hline
\multicolumn{10}{l}\footnotetext{all masses estimated using a temperature of 25 K; core A used 25 \& 300 K. }
    \end{tabular}
 \end{table*}

The continuum emission shows polarization at the level of 0.5 - 15 $\%$, exhibiting the well known polarization hole phenomenon, 
with the lowest polarization fractions occuring at the highest Stokes I intensities. Assuming that the polarization is due to
magnetically aligned dust grains, the derived B field orientations are shown in Fig. 2.  The orientation of the
polarization shows an ordered pattern, consistent with previous measurements (Cortes \& Crutcher, 2006) while reaching a
factor of 2 better resolution. It changes significantly over the map and can be decomposed into 
three regions corresponding to the dust peaks A, B-E and the much weaker C-D. The position angles are 
approximately perpendicular to each other between A and B-E. The statistics of the position angles are presented
in Table 2. The measurements include 50 detections with 3$\sigma$ or better signal to noise ratio and 17 with 2$\sigma$. 
The intrinsic dispersion of the position angle of the polarization, $\delta\phi$, is calculated by subtracting in quadrature 
an average position angle measurement error of 7.5 degree, arising from a 5 mJy rms on the Q and U images, from the observed 
rms of the position angles, $\sigma_{PA}$.

\begin{table*}
    \begin{tabular}{rrrrr}
      \multicolumn{5}{c}{Table 2: Polarization and Outflow Position Angles}\\
      \hline
      \hline

region&  PA$_{mean}$&  $\sigma_{PA}$&   $\delta\phi$ & Number of\\
&  deg.&  deg.&   deg & pixels\\

\hline
\hline
A  &   -39  &  14  &  12  & 13 \\             
B/E &     29  &  17   & 15 & 45  \\          
C/D  &      5 &    8 &    3 & 9  \\        
All &   13 &   31 &   30  & 67   \\       
large scale\footnotemark & 175 & $\pm$14 & & \\
OF-1 & 136 & & & \\
OF-2 & 135 & & & \\
OF-3 & 173 & & & \\

     \hline
\multicolumn{5}{l}\footnotetext{Dotson et al., 2010}
    \end{tabular}
 \end{table*}

For the region B-E, with the most numerous data points,  
we estimate the strength of the B-field using the Chandrasekhar-Fermi method. Following Crutcher et al (2004),
\\

$B_{pos} = 9.3 \times n(H_2)^{0.5}\delta V/\delta\phi$ \\

\noindent where $ B_{pos}, n(H_2), \delta V $ and $ \delta\phi$ are the plane-of-sky B-field ($\mu$G), density (cm$^{-3}$), FWHM of the
turbulent velocity dispersion (kms$^{-1}$) and the B-field position angle
dispersion (degree) respectively. We estimate $n(H_2)$ to be 10$^7$ cm$^{-3}$ by combining the masses of cores B and E and a size of
5$^{\prime\prime}$. For $\delta V $ we use a value of 3 kms$^{-1}$ from the single dish HCO$^+$ measurement of
Cortes (2010) with a 20$^{\prime\prime}$ beam, thus avoiding any interferometric spatial filtering. 
The CH$_3$CN linewidth, on a much smaller $\sim$ 1$^{\prime\prime}$ spatial scale, is still similar 
at 5 kms$^{-1}$ (section 4) and we consider the single dish measurement to be more representative of the velocity dispersion 
over the $\sim$ 7$^{\prime\prime}$  region over which the position angle dispersion is measured. The 
resulting $B_{pos}$ is 6 mG.  The mass-to-flux ratio estimated from these numbers is 3.5, and $\sim$ 1 when
a statistical geometric correction is applied, implying a critical to slightly super-critical condition. This is 
consistent with previous measurements (Cortes \& Crutcher 2006; Cortes, 2011) and would suggest that similar conditions 
are maintained on smaller spatial scales.  Nevertheless, we caution that these estimates are subject to large uncertainties.

\section {hot core} Strong emission from several hot-core species was detected towards core A.  Cores B and C also show 
emission at much weaker levels.  Here, we focus on the remarkable CH$_3$CN(J=19$ - $18) emission from core A, where 
11 K-ladder components were detected. The integrated emission and spectrum are shown in figures 1 \& 3 respectively.  
This data set only consists of the compact configuration observations and has a beam size of 2.1$''\times$1$''$. The 
emission is marginally resolved with a source size of 1.5$''\times$ 0.6$''$ (8000 $\times$ 3000 AU). 
The detection of a large number of lines from the K-ladder suggests a high temperature, the energy level for the 
K = 10 line being 885 K. The lowest K lines are optically thick, as seen from their
being about the same strength. Assuming all the K components are in LTE and trace the same gas, a grid 
search $\chi^2$ minimization was used to fit for temperature, column density and source size, including optical 
depth effects (Qiu \& Zhang 2009). Using the K = 2 to 10 components (fig. 3), we get 300 K, 2$\times$10$^{17}$ 
cm$^{-2}$ and 0.6$''$ (3000 AU) for the temperature, column density and source size. Fitting only 
for the optically thin high K components 7-10, the values are 420K, $4\times10^{16}$ cm$^{-2}$ and 1.2$''$. 
A systemic velocity of 101.5 kms$^{-1}$ was determined and the line width was set at 5 kms$^{-1}$ guided by gaussian fits to the 
K = 3,4 and 9 components, which appear to be free from blending, and the quality of the fit to the ladder. Such high 
temperatures are uncommon and only seen in a few other cases, some examples being 
Orion BN/KL ($\sim$ 400 K, Wilson et al 1993, Goddi et al 2011) and W51 IRS2 ($\sim$ 310 K, Mauersberger, 
Henkel \& Wilson, 1987). For comparison, the source size and luminosity can be used to obtain an
independent estimate of the temperature. 
Taking the luminosity from SED fits of 3 $\times$ 10$^4$ L$_\odot$ (Bally et al 2010; Herpin et al
2012), following Scoville \& Kwan (1976; also Scoville 2012), a dust temperature of 70-120 K (depening on
opacity) is obtained for a radial distance of 2500 AU (geometric mean of the semi major and minor axes), 
assuming heating by stellar radiation.  The Herpin et al model also shows a lower temperature ($\sim$ 150
K; as shown in their fig. 5) for the 2500 AU distance. If this dust temperature also characterizes CH$_3$CN 
excitation (Doty et al 2010), its disagreement with the measured temperature of 300-400K points to the possibility 
that stellar heating alone may not be sufficient to account for the hot CH$_3$CN. More detailed modeling and 
corroboration from other data will be needed for further clarity. In summary, we suggest that a high-mass proto-cluster 
may be forming here 
within a compact $\sim$ 0.01 pc radius region, although we do not yet have evidence for multiplicity. 

\newpage

\section {outflows} The outflows are mapped and studied using CO (3$-$2) emission. The CO data only consist of 
sub-compact array observations and thus have a poorer spatial resolution of 5$''$.
The CO line shows emission to large ($\sim \pm$ 70 kms$^{-1}$) velocities. In Fig 2, we 
present integrated emission over the velocity ranges 60$-$90 and 120$-$165 kms$^{-1}$, with the systemic
velocity based on our CH$_3$CN data being 101.5 kms$^{-1}$. The CO emission appears to trace three bipolar
outflows, two of which are associated with the cores C and F (OF-3 and OF-2 respectively) and the third is centered 
between core A and B (OF-1). Higher resolution observations will be needed to clarify the location of the driving source 
for OF-1.  Outflows 1 and 2 are oriented the same way but with opposed red and blue
lobes.  The the velocity ranges used were chosen to best delineate the outflows. We estimate outflow masses for the three 
outflows following the approach in Garden et al (1991). Assuming optically thin emission the outflow masses derived are 
12, 11 and 14 M$_\odot$  for OF-1,2 aad 3 respectively, for an excitation temperature 20 K. Based on the single dish CO(3-2) 
brightness temperature from Cortes et al (2011), 20 K was taken to be the lower limit and the mass estimate is not very
sensitive to temperature, staying within a factor of two for temperatures upto 180 K. The  massive outflows imply high-mass 
star formation. The outflows have time scales of $\sim$ 10$^4$ years, estimated using outflow extents and velocities  of 
5$^{\prime\prime}$ ($\sim$ 0.1 pc) and 25 kms$^{-1}$, pointing to their youth.

\section {Summary \& Discussion}

Our observations show multiple high-mass star formation in the W43-MM1 region. A compact hot core detected
in a number of spectral lines has a temperature of $\sim$ 400 K, derived from CH$_3$CN emission. This high
temperature 
can help in the investigation of high temperature chemistry probed by other spectral lines being studied (not 
included in this letter). Multiple massive outflows were mapped in CO.  Dust polarization measurements  over 
the main outflow show an alignment between the outflow and the B-field orientation on the plane of the sky. 

We now seek to compare our results with measurements of larger scale B-field reported by Dotson et al (2010).
These single dish measurements obtained with the CSO/Hertz have a beam size of $\sim$ 18$^{\prime\prime}$. While the 
the position angles for the 6 detections reported are spread over $\sim$ 90 deg, a range similar to our small 
scale measurements, the inner most four
measurements are well ordered, nearly parallel to each other. The outermost two are nearly perpendicular to
each other. We exclude the outer two measurements and take the average of the four closest measurements to
define the "large scale" field, oriented at a position angle of 85 deg. (175 deg. for polarization) with a
range of $\pm$14 deg. This "large scale" corresponds to $\sim$ 40$^{\prime\prime}$ ( $\sim$ 1 pc) and
the SMA measurements are on a $\sim$ 7$^{\prime\prime}$ ($\sim$ 0.1 pc) scale. It is possible to exclude only the
most deviant single dish measurement in which case the remaining five measurements trace a gradually changing
pattern with the inner most four measurements defining the larger scale field for comparison. As seen in
the figure, the field orientation over the main outflow (OF-1) is not parallel to this large
scale field measured from the single dish observations. This suggests the possibility that the magnetic
fields as seen on the small scales changed orientation from the nearly parallel distribution over the
immediate larger scale neighbourhood, either during the formation of the cores or the star formation
activity influenced the present morphology.

To assess the possibility of the outflow influencing the B-field morphology, we compare the total mechanical 
energy in the outflow with energy in the magnetic field, noting that the outflow extent and the region
over which  the B-field is mapped are comparable. We estimate the outflow energy to be $\sim$10$^{47}$ erg, 
($= Mv^2$), taking an outflowing mass $M$ of 10 M$_\odot$ with a velocity $v$ of 
25 kms$^{-1}$. The magnetic energy calculated is also $\sim$10$^{47}$ erg, considering the volume over which 
the B-field is measured to be a cylinder of length 
$l = $ 4$\times$10$^4$ AU (7$^{\prime\prime}$) and diameter 
2$\times$10$^4$ AU (4$^{\prime\prime}$), as seen in the map for the B-E region, and a magnetic energy density of $B^2/8\pi = $ 2 $\times$10$^{-6}$ 
erg.cm$^{-3}$, using the value of B from section 3.  As these energies are similar, based on these numbers 
it is not possible to say if the outflow influenced the orientation of the B-field. Given the uncertainties in 
these estimates, we would need a 
large difference to be able to make a more definitive statement. While the comparison is inconclusive, we 
note that although the outflow mapped in CO traces densities of $\sim$10$^3$ cm$^{-3}$ and the B-field is 
measured by polarization originating in a much denser medium ($\sim$10$^7$ cm$^{-3}$), it is still appropriate 
to compare the energies of the two. This is 
because, the outflow has cleared a lower density cavity in the dense medium into which it was originally driven 
and it could have influenced the magnetic field during its early phase. In addition, magnetic tension
can couple the magnetic fields in different regions. If alignment is caused by the outflow 
then the field strength cannot be estimated using the C-F method. However, the estimate in section 5 is still 
a measure of the upper limit to the B-field, as the position angle dispersion resulting from turbulence is 
reduced by the outflow induced alignment. Thus, the estimates above are to be only taken as indicative of the 
strength of the B-field. 

There are some studies of alignment between outflows and B-fields in the literature, but the results are 
inconclusive (see Li et al 2013 for a review and discussion; Curran \& Chrysostomou 2007,  Wolf et al 2003, 
Davidson et al 2011, Chapman et al 2013, Hull et al 2013). They are predominantly single dish observations 
towards low-mass star-forming regions,  except the Hull et al. study which used the CARMA interferometer. 
There is a clear conflict between the results of the most recent Chapman et al. and the Hull et al. studies, the
first showing alignment between the B-field and outflow orientations, and the second showing no correlation. As 
suggested by Chapman et al., (see also Li et al., 2013), the two may be reconciled by the fact that they trace 
different scales dominated by differrent 
processes. Our data correspond to the spatial scales of the Chapman et al study ($\sim$ 10000 AU) where an 
alignment is seen, as in our case. However, a simple global picture - a strong magnetic field on large scales 
directing collapse along its orientation leading to the formation of flattened pseudo-disk structures; rotation 
axes aligned to the B-field by magnetic breaking and alignment of outflows to this axis - is inconsistent with 
our observations. This is because (1) the field orientation in our map varies on small scales, and one of the three
outflows is not aligned to the other two; (2) the large scale field is not aligned to either of the directions
of the outflows or to the small scale B-field. Observations with finer spatial resolutions can help delineate potential 
pseudo-disk structures and study their rotation and relationship to the outflows and B-field orientations. The presence of 
multiple outflows within a small region presents a good opportunity to pursue this avenue.

Comments from an anonymous referee which helped improve the paper are gratefully acknowledged.

\begin{figure}
\plotone{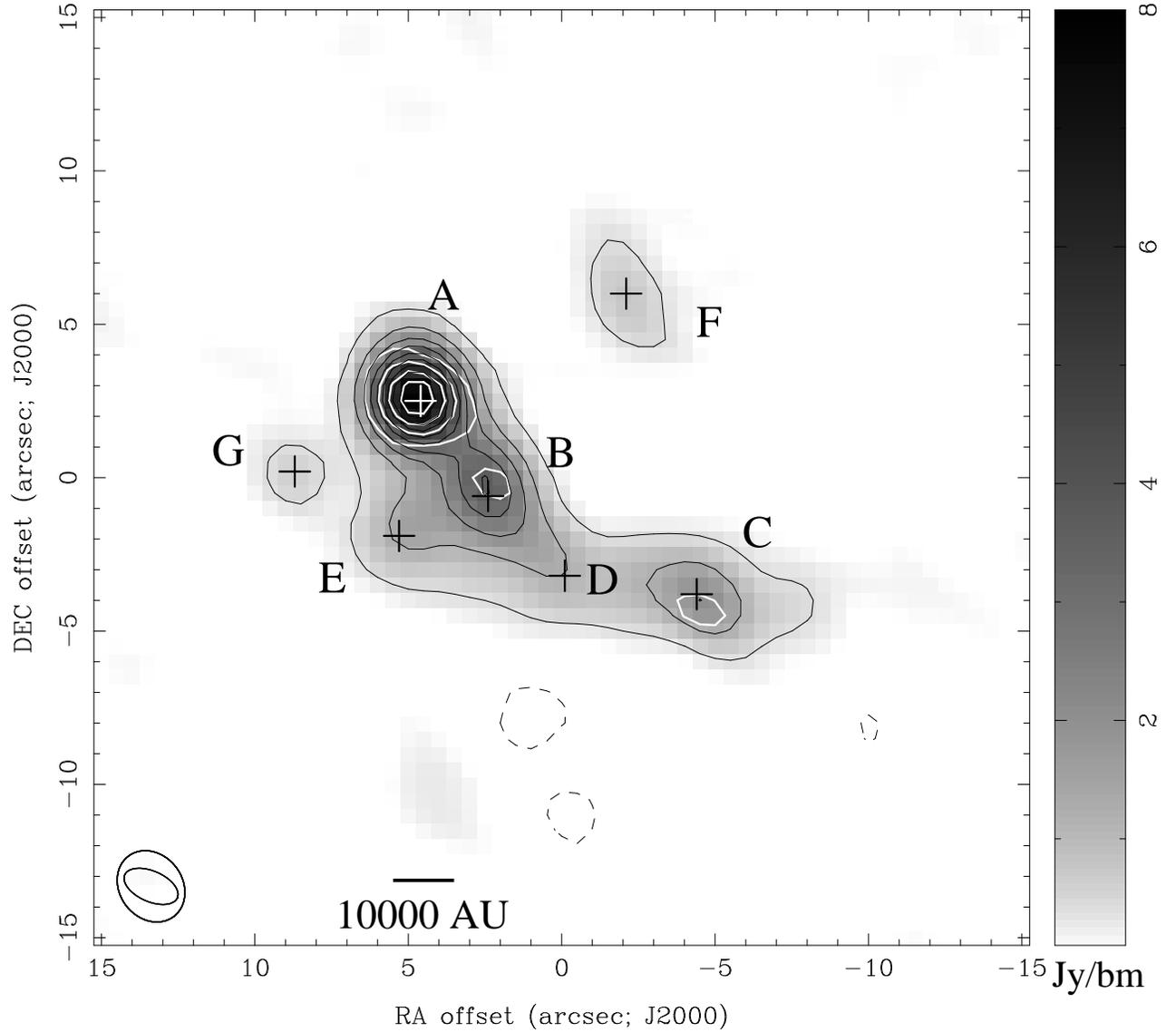}
\caption{The grey scale image and the black contours show the 345 GHz continuum emission. 
Continuum peaks brighter than the level of the highest negative contour were fitted and labeled A-G (listed in Table 1). 
The CH$_3$CN
emission integrated over the first four K-components is shown as white contours. The continuum contour
levels start at -0.4 Jy/beam with a step size of 0.8 Jy/beam, 1$\sigma$ being 0.1 Jy/beam. The line contours 
start at 37.5 Jy/beam-kms$^{-1}$ with a step size of 75 Jy/beam-kms$^{-1}$, 1$\sigma$ being 7 Jy/beam-kms$^{-1}$. 
The synthesized beams are shown at the bottom left (smaller beam - CH$_3$CN).}
\end{figure}

\begin{figure}
\plotone{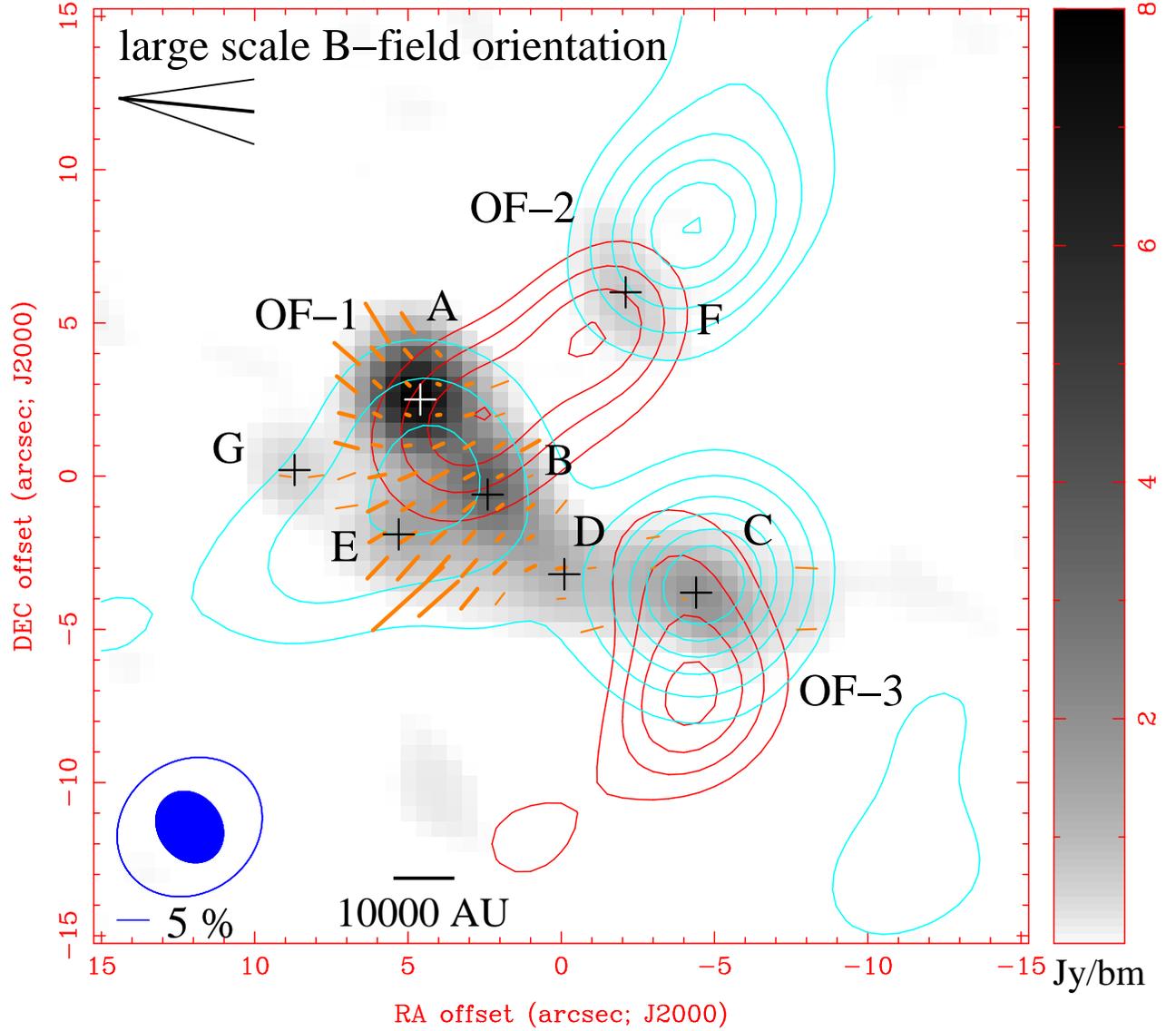}
\caption{Continuum emission, B-field and CO outflows. 
The CO contours start at 40 Jy/beam-kms$^{-1}$ and the
step size is 40 Jy/beam-kms$^{-1}$, 1$\sigma$ being 10 Jy/beam-kms$^{-1}$. The B-field orientations are shown as line segments with 
lengths proportional to fractional polarization. Thick and thin segments mark $>$ 3$\sigma$ and 2$\sigma$-3$\sigma$ measurements,
respectively. The orientation of the large scale B-field and its range are also shown. The synthesized
beams are shown at the bottom left (filled - continuum). A color version of this figure is available on-line.}
\end{figure}

\begin{figure}
\plotone{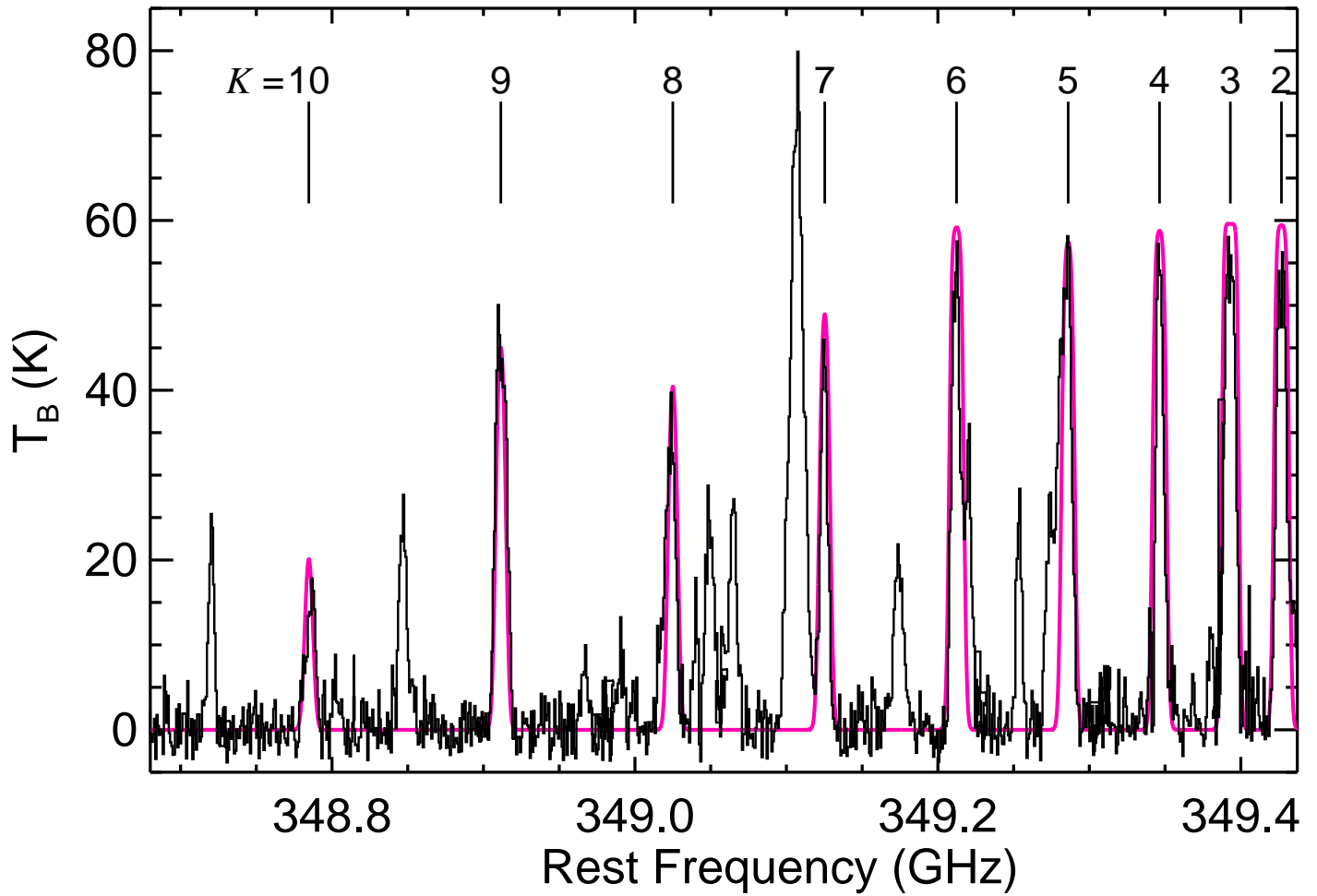}
\caption{Spectrum and fits for the CH$_3$CN (19-18), K=2 to K=10  lines. The K=0 and 1 components are
not shown. The spectrum was extracted over one synthesized beam at the peak. Other prominent lines seen in
this spectrum are tentatively identified to be from: CH$_3$OH (349.107 GHz), CH$_2$CHCN (348.991 GHz), HCOOCH$_3$ (349.048 \& 349.066 GHz), 
CH$_3^{13}$CN (349.253 \& 349.173 GHz) and C$_2$H$_5$OH (348.848 \& 348.720).}
\end{figure}

\newpage

\end{document}